# The critical O($N$) $\sigma$-model at dimension $2 < d < 4$: Hardy-Ramanujan distribution of quasi-primary fields and a collective fusion approach

K. Lang*and W. Rühl
Theoretische Physik,
Fachbereich Physik, Universität Kaiserslautern,
D-67653 Kaiserslautern, Germany

## Abstract

The distribution of qp-fields of fixed classes characterized by their O($N$) representations $Y$ and the number $p$ of vector fields from which they are composed at $N = \infty$ in dependence on their normal dimension $[\delta]$ is shown to obey a Hardy-Ramanujan law

$$A\,[\delta]^{-\alpha}\,\exp\left(B\,[\delta]^{\frac{1}{2}}\right)$$

at leading order in a $\frac{1}{N}$-expansion. We develop a method of collective fusion of the fundamental fields which yields arbitrary qp-fields and resolves any degeneracy.

## 1 Introduction

In a series of papers analysing the critical O($N$) $\sigma$-model [1-8], we have been able to classify quasi-primary fields ("qp-fields") by ordering them into classes $(Y, p)$. $Y$ is the Young frame of the irreducible O($N$) representation and $p$ is the number of vector fields $\vec{S}$ which in the $N = \infty$ limit are involved in the qp-field. The class $(\emptyset, 0)$ corresponds to the $\alpha$-field theory which is closed under operator product expansions. Classes $(Y, p)$ describe irreducible modules with respect to this $\alpha$-field subalgebra.

In the action of the O($N$)-vector $\sigma$-model

$$\mathcal{Z} = \int D[\vec{S}]D[\alpha] \exp\left\{-\int dx \left[\frac{1}{2}(\partial_\mu \vec{S})^2(x) + z^{\frac{1}{2}}\alpha(x)(\vec{S}^2(x) - 1)\right]\right\} \quad (1.1)$$

we have two fundamental fields: the O($N$) vector field $\vec{S}$ and the auxiliary (or Lagrange multiplier) field $\alpha$. All qp-fields are derived from $\vec{S}$ and $\alpha$ by multiple operator product expansions (OPE). We have made two observations.

---

*e-mail: lang@gypsy.physik.uni-kl.de

1. At $N = \infty$ all qp-fields can be obtained already from a suitably chosen multiple OPE in the form of a normal product of $p$ fields $\vec{S}$ and $r$ fields $\alpha$. The class $(Y, p)$ contains qp-fields with variable $r$.

2. There is a combinatorial analogy between bound states of $p$ hypothetical particles $\vec{S}$ and $r$ particles $\alpha$ and the qp-fields with the same parameters $p$ and $r$.

Both observations will be amalgamated into a combinatorial approach giving all qp-fields at $\mathcal{O}(1)$ and their anomalous dimensions at $\mathcal{O}(\frac{1}{N})$ but no fusion constants for fusion of two qp-fields into a third qp-field. Instead all $r + p$ fundamental fields participating are fused simultaneously into one qp-field ("collective fusion").

In section 2 we give some background information on Hardy-Ramanujan distributions and the combinatorics of bound-state models which underly our approach. In section 3 we prove the Hardy-Ramanujan distribution for the class $(\emptyset, 0)$. The proof is constructive and can be used to derive all qp-fields at $\mathcal{O}(1)$. The distinction between qp-fields and derivative fields is studied in section 4. We construct an orthogonal projection which decomposes any field into a qp-field and and a derivative field component. This new method is tested for its applicability in section 5. We analyze a simple but nontrivial example that is not accesssible with our earlier approach: the twofold degenerate level at $[\delta] = 12$, $l = 6$ in the class $(\emptyset, 0)$. We have to fuse three $\alpha$-fields collectively. Stepwise fusion with an intermediate qp-field with $[\delta] = 4$, $l = 0$ did not resolve the degeneracy. In section 5 we consider the generalization of this approach to arbitrary classes. We do not work it out completely since developing a computer program for the construction (at $\mathcal{O}(1)$ ) and identification (at $\mathcal{O}(\frac{1}{N})$ ) of any qp-field appears now feasible.

## 2 The Verma module of the Virasoro algebra and a bound state model

The Virasoro algebra spanned by the basis

$$\{C, L_0, L_{\pm 1}, L_{\pm 2}, \ldots\} \tag{2.1}$$

with commutation relations

$$[L_n, L_m] = (n - m)L_{n+m} + \frac{1}{12}C\delta_{n+m,0}n(n^2 - 1) \tag{2.2}$$

possesses highest weight representations $\mathcal{V}(c, h)$ which are cyclically generated from a highest weight state $\phi_{c,h}$

$$L_0|\phi_{c,h}\rangle = h|\phi_{c,h}\rangle, \qquad C|\phi_{c,h}\rangle = c|\phi_{c,h}\rangle, \qquad L_n|\phi_{c,h}\rangle = 0, \quad n > 0 \tag{2.3}$$

so that $\mathcal{V}(c, h)$ is spanned by states

$$\phi_{\{n_1, n_2, \ldots, n_k\}} = L_{-n_1} L_{-n_2} L_{-n_3} \ldots L_{-n_k} |\phi_{c,h}\rangle, \qquad n_1 \geq n_2 \geq n_3 \geq \ldots \geq n_k > 0 \tag{2.4}$$

A degree is obtained by

$$deg\phi_{\{n_1, n_2, \ldots, n_k\}} = \sum_{i=1}^{k} n_i = n \tag{2.5}$$

We consider the generic case in which all states $\phi$ are independent.

The subspace $\mathcal{V}_n$ of $\mathcal{V}(c,h)$ of vectors of common degree $n$ has dimension

$$dim\, \mathcal{V}_n \;=\; p(n) \tag{2.6}$$

where $p(n)$ is the partition function defined by the generating function

$$\sum_{n=0}^{\infty} p(n)\, x^n \;=\; \Big[\prod_{k=1}^{\infty} (1-x^k)\Big]^{-1} \tag{2.7}$$

For $n \to \infty$ we have the Hardy-Ramanujan formula [9]

$$p(n) \;=\; \frac{1}{4\sqrt{3}n} \exp\Big(\pi\sqrt{\frac{2}{3}n}\Big) \Big(1 + \mathcal{O}\Big(\frac{\log n}{n^{\frac{1}{4}}}\Big)\Big) \tag{2.8}$$

It is well known [10] that each state $\phi$ can be identified with a "chiral" qp-field $\varphi(z)$ of dimension $h+n$ or a derivative of a lower degree qp-field. The derivative fields arise by the identification

$$L_{-1} \;\leftrightarrow\; \frac{\partial}{\partial z} \tag{2.9}$$

A qp-field is determined by a simple transformation behaviour under the conformal group, which in this case is the Lie group generated from $\{L_0, L_{\pm 1}\}$.

By an elementary argument one can show that $\mathcal{V}_n$ is spanned by $p(n-1)$ derivative fields and

$$\Delta(n) \;=\; p(n) - p(n-1) \tag{2.10}$$

qp-fields. From (2.8) we obtain

$$\Delta(n) \;=\; \frac{\pi}{12\sqrt{2}}\, n^{-\frac{3}{2}} \exp\Big(\pi\sqrt{\frac{2}{3}n}\Big) \Big(1 + \mathcal{O}\Big(\frac{\log n}{n^{\frac{1}{4}}}\Big)\Big) \tag{2.11}$$

The distribution (2.10) reappears in a completely different context. Take a system of $r$ $\alpha$-particles (helium nuclei) considered as elementary particles. We separate the motion of the center of mass in the wave function

$$\exp(iK \cdot X)\, \Psi(x_1, x_2, \ldots, x_r), \qquad x_i \in I\!\!R_3, \qquad X \;=\; \sum_{i=1}^{r} x_i \tag{2.12}$$

Then the remaining internal wave function is translation invariant

$$E(a)\, \Psi \;=\; \Psi, \qquad \forall a \in I\!\!R_3 \tag{2.13}$$

with

$$E(a) \;=\; \exp\big\{a E'(0)\big\} \tag{2.14}$$

$$E'(0) \;=\; \sum_{i=1}^{r} \frac{\partial}{\partial x_i} \tag{2.15}$$

Moreover, $\Psi$ is bose symmetric (under $x_i \leftrightarrow x_j$).

If $\Psi$ corresponds to a state of total (orbital) angular momentum $L$ then it should factor into a solid spherical harmonic of degree $L$ and a rotation invariant function. Such solid spherical harmonics are constructed as traceless symmetric tensors $Y^L$. We contract one with an arbitrary vector $q$ and impose Bose symmetry

$$(q)^L_\otimes Y^L_{\{l_1,l_2,\ldots,l_r\}}(x_1, x_2, \ldots, x_r) = \sum_{\substack{\text{permutations} \\ \text{of } \{x_i\}_{i=1}^r}} \prod_{i=1}^r (q \cdot x_i)^{l_i} - \text{trace terms} \qquad (2.16)$$

Knowledge of the trace terms which render $Y^L$ traceless is not necessary. Of course it can be assumed that the $\{l_i\}_{i=1}^r$ are ordered

$$l_1 \geq l_2 \geq l_3 \geq \ldots \geq l_r \geq 0, \qquad \sum_{i=1}^r l_i = L \qquad (2.17)$$

so that any such solid spherical harmonics is characterized by partitions of L <u>into at most r parts</u>, the dimension of the space $V_L^{(r)}$ therefore being equal to the number of these partitions

$$dim\, V_L^{(r)} = p_r(L), \qquad (p_r(L) = p(L), \quad r \geq L) \qquad (2.18)$$

We still have to impose the constraint (2.13). We define linear combinations

$$\mathcal{Y} = \sum_{\substack{\text{partitions of } L \\ \text{of length} \leq r}} C_{\{l_1,l_2,\ldots,l_r\}} (q)^L_\otimes Y^L_{\{l_1,l_2,\ldots,l_r\}}(x_1, \ldots, x_r) \qquad (2.19)$$

The constraint of translation invariance (2.13) amounts then by (2.14) to

$$E'(0)\, \mathcal{Y} = 0 \qquad (2.20)$$

or

$$\sum_{\substack{\text{partitions of } L \\ \text{of length} \leq r}} \sum_{i=1}^r l_i\, C_{\{l_1,l_2,\ldots,l_r\}} (q)^{L-1}_\otimes Y^{L-1}_{\Omega\{l_1,l_2,\ldots,l_i-1,\ldots,l_r\}}(x_1, \ldots, x_r) = 0 \qquad (2.21)$$

where $\Omega$ reorders the sequence

$$l_1, l_2, \ldots, l_i - 1, \ldots, l_r \qquad (2.22)$$

Let

$$K_L^{(r)} = ker\, E'(0) \cap V_L^{(r)} \qquad (2.23)$$

This has the dimension of $V_L^{(r)}$ minus the number of (linear independent) constraints for the $C_{\{\ldots\}}$ in (2.21)

$$dim\, K_L^{(r)} = p_r(L) - p_r(L-1) \qquad (2.24)$$

For numerical calculations Euler's recursion relation for the $p_r(n)$

$$p_r(n) = p_{r-1}(n) + p_r(n-r), \qquad p_0(n) = 0 \qquad (2.25)$$

is of great value.

## 3   Conformal field theory at $d > 2$

The field $\alpha$ in the O($N$) vector $\sigma$-model creates a class $(Y, p)$ of qp-fields with

$$Y = \emptyset, \qquad p = 0 \qquad (3.1)$$

This class forms a closed set under OPE and creates the smallest field subalgebra (a Wightman field theory) of the O($N$) vector $\sigma$-model. The qp-fields can be ascribed to levels of a spectral diagram (Fig. 1). Each level may be multiply (finitely) occupied. A level is labelled by

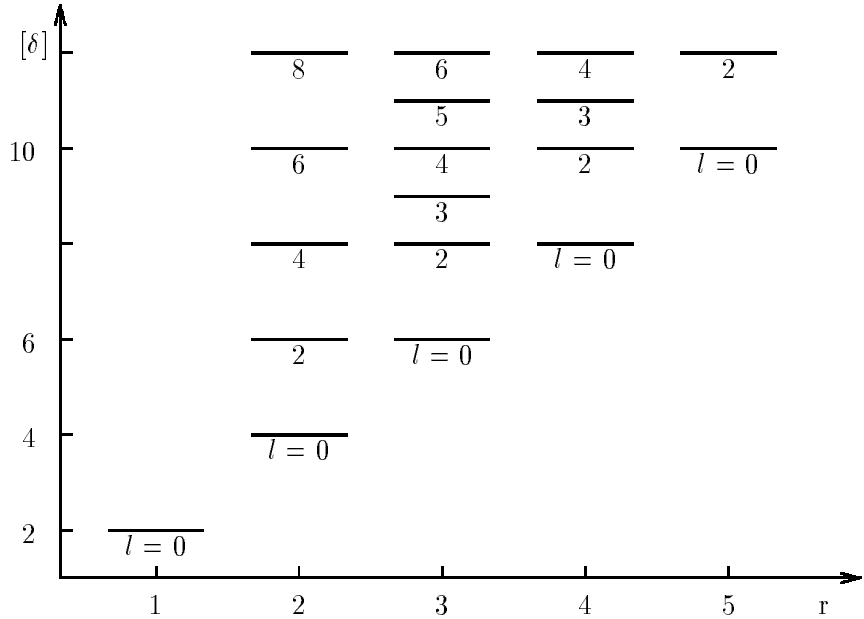

Figure 1: The spectral diagram of the class $(\emptyset, 0)$

1. the normal dimension $[\delta]$ which is the field dimension $\delta$ in the $N = \infty$ limit.

2. the number $r$ of $\alpha$ fields of which the field is composed at $N = \infty$

3. the tensor rank $l$ of the field, which is a symmetric traceless tensor field.

The occupancy number of a level is called its degeneracy. At $\mathcal{O}(\frac{1}{N})$, the degeneray is lifted due to appearance of different anomalous dimensions - allowing to identify qp-fields.

By a stepwise fusion (i.e. OPE) forcing each intermediate qp-field to be scalar and finally applying conformal exchange amplitudes (that is: harmonic analysis of the conformal group) we have created qp-fields in this class up to $l = 6$ and resolved their anomalous dimension [7]. We have found that degeneracy sets in at $r \geq 4, l \geq 4$. But we suspected that the restriction to scalar intermediate fields makes it impossible to find all qp-fields for large $l$. In fact, the approach we are developing here yields modifications to these degeneracies for $r \geq 3, l \geq 6$, but this time we are sure to have a handle to produce all qp-fields.

In an OPE of three scalar fields

$$A(x) B(y) = \left((x-y)^2\right)^{\frac{1}{2}(\delta_C - \delta_A - \delta_B)} C(y) + \ldots \tag{3.2}$$

we can, in the limit $N = \infty$, denote $C$ as the normal product of $A$ and $B$

$$C(y) = : A(y) B(y) : \tag{3.3}$$

if

$$\lim_{N \to \infty} (\delta_C - \delta_A - \delta_B) = 0 \tag{3.4}$$

It's easy to derive Leibniz's formula

$$(q \cdot \partial)^n C(y) = \sum_{k=0}^{n} \binom{n}{k} : (q \cdot \partial)^{n-k} A(y) (q \cdot \partial)^k B(y) : \tag{3.5}$$

by differentiating (3.2) with respect to a translation ($q$ is an arbitrary constant vector).

We define a traceless symmetric tensor field $\phi_{\{l_1, l_2, \ldots, l_r\}}$ from $r$ fields $\alpha$ by

$$(q)_{\otimes}^L \phi_{\{l_1, l_2, \ldots, l_r\}}(y) = : (q \cdot \partial)^{l_1} \alpha(y) (q \cdot \partial)^{l_2} \alpha(y) \ldots (q \cdot \partial)^{l_r} \alpha(y) : - \text{trace terms} \tag{3.6}$$

where we assume, without restriction of generality, again validity of (2.17). A qp-field of tensor rank $L$ must then be a linear combination of such fields

$$\sum_{\substack{\text{partitions of } L \\ \text{of length } \leq r}} C_{\{l_1, l_2, \ldots, l_r\}} (q)_{\otimes}^L \phi_{\{l_1, l_2, \ldots, l_r\}}(y) \tag{3.7}$$

A qp-field has a simple transformation behaviour under conformal inversions

$$I : y \to \frac{-y}{y^2} =: \eta \tag{3.8}$$

Namely, each tensor index transforms as

$$T_I \phi_\mu(y) = (y^2)^{-\delta_\phi} r_{\mu\nu}(y) \phi_\nu(\eta) \tag{3.9}$$

with

$$r_{\mu\nu}(y) = 2\frac{y_\mu y_\nu}{y^2} - g_{\mu\nu} \qquad (3.10)$$

From (3.6) we get instead

$$T_I(q)_\otimes^L \phi_{\{l_1,l_2,\ldots,l_r\}}(y) = (y^2)^{-r\delta_2} : \left(q \cdot Q(\eta)\right)^{l_1} \alpha(\eta) \cdots \left(q \cdot Q(\eta)\right)^{l_r} \alpha(\eta) :$$
$$- \text{ trace terms}$$

where

$$Q_\mu(\eta) = \eta^2 r(\eta)_{\mu\nu}\frac{\partial}{\partial \eta_\nu} + 2\delta_2 \eta_\mu \qquad (3.11)$$

and

$$\delta_2 = dim\, \alpha = 2 + \mathcal{O}(\tfrac{1}{N}) \qquad (3.12)$$

Now we bring all $\frac{\partial}{\partial \eta}$ to the right of all multiplications with $\eta$

$$\left(q \cdot Q(\eta)\right)^l = \sum_{k=0}^{l} A_k^l (q \cdot \eta)^k \left(\eta^2\, q \cdot r(\eta)\right)^{l-k} \cdot (\partial)^{l-k} \qquad (3.13)$$

By recursion one can prove that

$$A_k^l = 2^k \binom{l}{k} (\delta_2 + l - k)_k \qquad (3.14)$$

Only the terms $k = 0$ lead to the correct transformation behaviour (3.9) with

$$\delta_\phi = r\,\delta_2 + L \qquad (3.15)$$

Therefore all other terms must cancel if $T_I$ is applied to (3.7). How many constraints are this?
Consider a space of polynomials $V$ of arbitrary degree and define the operator

$$E(a)\, x^l = \sum_{k=0}^{l} A_k^l\, a^k\, x^{l-k} \qquad (3.16)$$

with $A_k^l$ from (3.14). It is then easy to show that in this space

$$E(a)\, E(b) = E(a + b) \qquad (3.17)$$

implying, as in (2.14)

$$E(a) = \exp\left(a\, E'(0)\right) \qquad (3.18)$$

where

$$E'(0) x^l = A_1^l\, x^{l-1} \qquad (3.19)$$
$$= 2l(\delta_2 + l - 1)\, x^{l-1} \qquad (3.20)$$

We can extend the operation $E(a)$ to an $r$-fold tensor product $V_\otimes^r$:

$$\tilde{E}(a)\left(x^{l_1}, x^{l_2}, \ldots, x^{l_r}\right) = \left(E(a)x^{l_1}, \ldots, E(a)x^{l_r}\right) \qquad (3.21)$$

(no ordering of the $l_i$ of course) so that $\tilde{E}(a)$ satisfies (3.17), (3.18) too. The algebraic structure of (3.16), (3.21) is analogous to (3.13), (3.11).

Let us therefore define the vector space $V_L^{(r)}$ of elements (3.7) and

$$K_L^{(r)} = ker\,\tilde{E}'(0) \cap V_L^{(r)} \qquad (3.22)$$

Then $K_L^{(r)}$ has elements satisfying

$$\sum_{\substack{\text{partitions of } L \\ \text{of length} \leq r}} C_{\{l_1,l_2,\ldots,l_r\}} \sum_{i=1}^{r} l_i(\delta + l_i - 1)(q)_\otimes^{L-1} \phi_{\Omega\{l_1,\ldots,l_i-1,\ldots,l_r\}}(y) = 0 \qquad (3.23)$$

where $\Omega$ as in (2.21) defines a reordering. Since (2.18) is valid, (2.24) follows as well.

Now we consider the whole class $(\emptyset, 0)$ (Fig. 1). The number of qp-fields for fixed

$$[\delta] = n \qquad (3.24)$$

is

$$dim(n) = \sum_{r=1}^{[\frac{n}{2}]} \left(p_r(n - 2r) - p_r(n - 2r - 1)\right) \qquad (3.25)$$

A fit to $dim(n)$ for $100 \leq n \leq 1000$ by the ansatz

$$\log dim(n) = \pi\sqrt{\frac{2}{3}n} - 2\log n + \sum_{s=0}^{N} \frac{a_s}{n^{\frac{s}{2}}} \qquad (3.26)$$

showed an excellent agreement to a Hardy-Ramanujan distribution law. The coefficients for $N = 2$ were

$$a_0 = -1.4382 \quad (=\log A) \qquad (3.27)$$
$$a_1 = -3.659 \qquad (3.28)$$
$$a_2 = +1.47 \qquad (3.29)$$

In Table 1 we list the function $dim(n)$ with the number $dst(n)$, the number of qp-fields found by stepwise fusion with all intermediate states restricted to be scalar [7]. The deviation arises first for $n = 12$ and increases strongly afterwards.

In Table 2 we plot the correct degeneracy function

$$mult(\emptyset, 0; 2r + l, l) = p_r(l) - p_r(l - 1) \qquad (3.30)$$

| $n$ | 2 | 3 | 4 | 5 | 6 | 7 | 8 | 9 | 10 | 11 | 12 | 13 | 14 | 15 | 16 | 17 |
|---|---|---|---|---|---|---|---|---|---|---|---|---|---|---|---|---|
| $dim(n)$ | 1 | 0 | 1 | 0 | 2 | 0 | 3 | 1 | 4 | 2 | 7 | 3 | 10 | 7 | 14 | 11 |
| $dst(n)$ | 1 | 0 | 1 | 0 | 2 | 0 | 3 | 1 | 4 | 2 | 6 | 3 | 8 | 5 | 10 | ... |

Table 1: The number $dim(n)$ of qp-fields for given normal dimension $[\delta] = n$ as compared with the number $dst(n)$ found by the stepwise fusion procedure.

| $l$ | | | | | | | | | | | | |
|---|---|---|---|---|---|---|---|---|---|---|---|---|
| 12 | 0 | 1 | 3 | 7 | 10 | 14 | 16 | 18 | 19 | 20 | 20 | 21 |
| 11 | 0 | 0 | 2 | 4 | 7 | 9 | 11 | 12 | 13 | 13 | 14 | 14 |
| 10 | 0 | 1 | 2 | 5 | 7 | 9 | 10 | 11 | 11 | 12 | 12 | 12 |
| 9 | 0 | 0 | 2 | 3 | 5 | 6 | 7 | 7 | 8 | 8 | 8 | 8 |
| 8 | 0 | 1 | 2 | 4 | 5 | 6 | 6 | 7 | 7 | 7 | 7 | 7 |
| 7 | 0 | 0 | 1 | 2 | 3 | 3 | 4 | 4 | 4 | 4 | 4 | 4 |
| 6 | 0 | 1 | 2 | 3 | 3 | 4 | 4 | 4 | 4 | 4 | 4 | 4 |
| 5 | 0 | 0 | 1 | 1 | 2 | 2 | 2 | 2 | 2 | 2 | 2 | 2 |
| 4 | 0 | 1 | 1 | 2 | 2 | 2 | 2 | 2 | 2 | 2 | 2 | 2 |
| 3 | 0 | 0 | 1 | 1 | 1 | 1 | 1 | 1 | 1 | 1 | 1 | 1 |
| 2 | 0 | 1 | 1 | 1 | 1 | 1 | 1 | 1 | 1 | 1 | 1 | 1 |
| 1 | 0 | 0 | 0 | 0 | 0 | 0 | 0 | 0 | 0 | 0 | 0 | 0 |
| 0 | 1 | 1 | 1 | 1 | 1 | 1 | 1 | 1 | 1 | 1 | 1 | 1 |
| | 1 | 2 | 3 | 4 | 5 | 6 | 7 | 8 | 9 | 10 | 11 | 12 | $r$ |

Table 2: Correct degeneracy of the levels $(\emptyset, 0; 2r+l, l)$ ($[\delta] = 2r+l$) found with the collective fusion procedure. To the right of the stairway curve the degeneracy is independent of $r$.

whereas in Table 3 we repeat the function [7]

$$mult(\emptyset, 0; 2r+l, l) = \min\left\{N_{rl}, [\frac{l}{2}]\right\} \qquad (3.31)$$

$$N_{rl} = \begin{cases} \frac{1}{2}r - 1 & : \quad r \text{ even}, l \text{ odd} \\ [\frac{r}{2}] & : \quad r \text{ odd or } l \text{ even} \end{cases}$$

now known to deviate (and thus to be too small) for $l \geq 6$.

## 4  Quasiprimary versus derivative fields

The spaces $V_L^{(r)}$ which are spanned by the normal products of $r$ fields $\alpha$ (3.6)

$$\phi_{\{l_1,\ldots,l_r\}}(0), \qquad \sum_{i=1}^r l_i = L \qquad (4.1)$$

| $l$ | | | | | | | | | | | | |
|---|---|---|---|---|---|---|---|---|---|---|---|---|
| 12 | 0 | 1 | 1 | 2 | 2 | 3 | 3 | 4 | 4 | 5 | 5 | 6 |
| 11 | 0 | 0 | 1 | 1 | 2 | 2 | 3 | 3 | 4 | 4 | 5 | 5 |
| 10 | 0 | 1 | 1 | 2 | 2 | 3 | 3 | 4 | 4 | 5 | 5 | 5 |
| 9 | 0 | 0 | 1 | 1 | 2 | 2 | 3 | 3 | 4 | 4 | 4 | 4 |
| 8 | 0 | 1 | 1 | 2 | 2 | 3 | 3 | 4 | 4 | 4 | 4 | 4 |
| 7 | 0 | 0 | 1 | 1 | 2 | 2 | 3 | 3 | 3 | 3 | 3 | 3 |
| 6 | 0 | 1 | 1 | 2 | 2 | 3 | 3 | 3 | 3 | 3 | 3 | 3 |
| 5 | 0 | 0 | 1 | 1 | 2 | 2 | 2 | 2 | 2 | 2 | 2 | 2 |
| 4 | 0 | 1 | 1 | 2 | 2 | 2 | 2 | 2 | 2 | 2 | 2 | 2 |
| 3 | 0 | 0 | 1 | 1 | 1 | 1 | 1 | 1 | 1 | 1 | 1 | 1 |
| 2 | 0 | 1 | 1 | 1 | 1 | 1 | 1 | 1 | 1 | 1 | 1 | 1 |
| 1 | 0 | 0 | 0 | 0 | 0 | 0 | 0 | 0 | 0 | 0 | 0 | 0 |
| 0 | 1 | 1 | 1 | 1 | 1 | 1 | 1 | 1 | 1 | 1 | 1 | 1 |
|   | 1 | 2 | 3 | 4 | 5 | 6 | 7 | 8 | 9 | 10 | 11 | 12 $r$ |

Table 3: Incomplete degeneracy of the levels $(\emptyset, 0; 2r+l, l)$ ($[\delta] = 2r+l$) found according to the stepwise fusion procedure [7].

can be combined in one big space by the orthogonal direct sum

$$V^{(r)} = \bigoplus_{L=0}^{\infty} V_L^{(r)} \qquad (4.2)$$

Then we consider two operators acting in $V^{(r)}$.

First we study (3.23)

$$D_r = \tilde{E}'(0) \qquad (4.3)$$

$$D_r \phi_{\{l_1,\ldots,l_r\}} = \sum_i l_i(\delta + l_i - 1)\phi_{\Omega\{l_1,\ldots,l_i-1,\ldots,l_r\}} \qquad (4.4)$$

The qp-fields correspond to the kernels of $D_r$ in $V_L^{(r)}$ : $K_L^{(r)} \subset V_L^{(r)}$. Deriving a normal product and applying the Leibniz rule (3.5) leads us to introduce the derivative operator $\Delta_r$

$$\Delta_r \phi_{\{l_1,\ldots,l_r\}} = \sum_i \phi_{\Omega\{l_1,\ldots,l_i+1,\ldots,l_r\}} \qquad (4.5)$$

We denote the image of $V_L^{(r)}$ in $V_{L+1}^{(r)}$ by $H_{L+1}^{(r)}$

$$\Delta_r V_L^{(r)} = H_{L+1}^{(r)} \subset V_{L+1}^{(r)} \qquad (4.6)$$

Elements of $H_{L+1}^{(r)}$ are called "derivative fields".

The main result is now that (Fig. 2)

$$V_L^{(r)} = K_L^{(r)} \oplus H_L^{(r)} \qquad (4.7)$$

so that any vector of $V_L^{(r)}$ can be decomposed uniquely

$$\psi \in V_L^{(r)}$$
$$\psi = \psi_0 + \psi_1, \quad \psi_0 \in K_L^{(r)}, \quad \psi_1 \in H_L^{(r)} \tag{4.8}$$

We need to prove this assertion in a constructive fashion which permits us to calculate $\psi_0$ and $\psi_1$.

First assume that a decomposition exist. Then

$$\psi_1 = \Delta_r \chi, \quad \chi \in V_{L-1}^{(r)} \tag{4.9}$$
$$D_r \psi = D_r \Delta_r \chi \tag{4.10}$$

We denote the restriction of $D_r \Delta_r$ to $V_{L-1}^{(r)}$ by $d_{L-1}^{(r)}$

$$d_{L-1}^{(r)} = D_r \Delta_r P_{L-1}^{(r)} \tag{4.11}$$

where $P_{L-1}^{(r)}$ is the projection of $V^{(r)}$ on $V_{L-1}^{(r)}$ by (4.1). If $d_{L-1}^{(r)}$ is nonsingular then by (4.10)

$$\chi = \left(d_{L-1}^{(r)}\right)^{-1} D_r \psi \tag{4.12}$$

and

$$\psi_1 = \Delta_r \left(d_{L-1}^{(r)}\right)^{-1} D_r \psi \tag{4.13}$$
$$\psi_0 = \psi - \psi_1 \tag{4.14}$$

On the other hand, assuming $d_{L-1}^{(r)}$ to be nonsingular, we can define $\psi_0$ and $\psi_1$ for each $\psi$ by (4.13), (4.14). Then $\psi_1 \in H_L^{(r)}$ trivially, and

$$\begin{aligned} D_r \psi_0 &= D_r \psi - D_r \psi_1 \\ &= D_r \psi - \left(D_r \Delta_r\right)\left(d_{L-1}^{(r)}\right)^{-1} D_r \psi \\ &= 0 \end{aligned} \tag{4.15}$$

so that also $\psi_0 \in K_L^{(r)}$.

So our assertion (4.7), (4.8) is proved, provided all $d_L^{(r)}$ are nonsingular. The matrices of

$$D_r P_L^{(r)}, \quad \Delta_r P_{L-1}^{(r)} \tag{4.16}$$

are easily calculated with (??), (4.4), (4.5). The matrix $\left(d_L^{(r)}\right)$ of $d_L^{(r)}$ is then

$$\left(d_L^{(r)}\right) = \left(D_r P^{(r)}\right)\left(\Delta_r P_{L-1}^{(r)}\right) \tag{4.17}$$

and we found (provided the partitions are in lexicographical order)

$$det\left(d_L^{(r)}\right) = \prod_{n=0}^{L} \left((r\delta + L + n)(L + 1 - n)\right)^{dim(K_n^{(r)})} \tag{4.18}$$

with $dim(K_n^{(r)})$ from (2.24).

Only for negative values of the field dimension (which are excluded by physical arguments) the operator $d_L^{(r)}$ becomes singular. The proof of (4.18) is elementary but lengthy and therefore we skip it here.

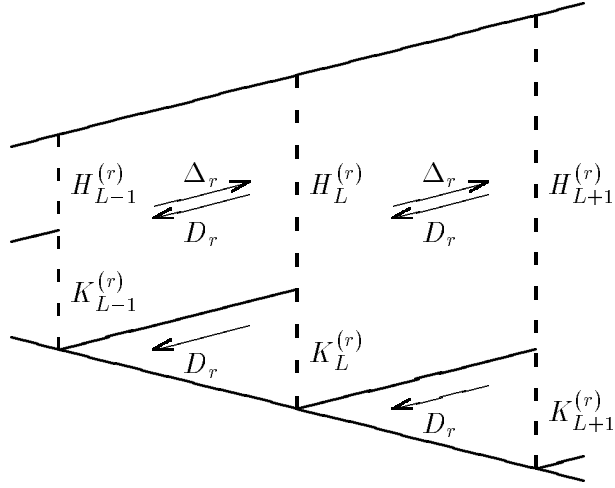

Figure 2: The subspaces $K_L^{(r)}$ and $H_L^{(r)}$ and the actions of the operators $D_r$ and $\Delta_r$.

## 5  An example for a degenerate level

According to Table 2 the level

$$C_l^r \quad : \quad (\emptyset, 0; 2r + l, l), \qquad r = 3, \quad l = 6 \tag{5.1}$$

has degeneracy two, which has not been resolved before. We study its resolution in full detail as a test for the "collective fusion algorithm".

However, we begin with an application of the "stepwise fusion algorithm", corresponding to the process depicted in Fig. 3 ($r = 3$, but $l$ arbitrary). The fusion coefficients for the final fusion step

$$\alpha \otimes \alpha_\otimes^2 \quad \to \quad C_l^3 \tag{5.2}$$

are

$$\gamma_l \;=\; \frac{(\delta_2)_l}{3(3\delta_2 + l - 1)_l}\Big\{2(\delta_2)_l + (-1)^l (2\delta_2)_l\Big\} \tag{5.3}$$

obtained from [7], (2.21) by inserting

$$\sigma_m \;=\; \frac{1}{3}\delta_{m0} + \frac{2}{3}(\delta_2)_m \tag{5.4}$$

$$\delta_{A'} \;=\; \delta_A \;=\; \delta_2, \quad \delta_B \;=\; 2\delta_2 \tag{5.5}$$

We set $\delta_2 = 2$ at the end.

The anomalous dimensions are obtained from [7], (2.23) with

$$\varphi_m = \frac{2}{3}\phi_m^{(2)} + \frac{1}{3}(\delta_2)_m \phi_0^{(2)} \tag{5.6}$$

and ([7], (3.17))

$$\phi_m^{(2)} = \frac{4\mu(2\mu-3)}{\mu-2}\frac{m!}{(\mu)_m}\left\{(2\mu-3)\left[(\mu-1)_m + \frac{(\mu-1)_{m+1}-(m+1)!}{\mu-2}\right] \right. \\ \left. -(2\mu-5)\frac{(\mu-2)_{m+1}-(m+1)!}{\mu-3}\right\} \tag{5.7}$$

Using

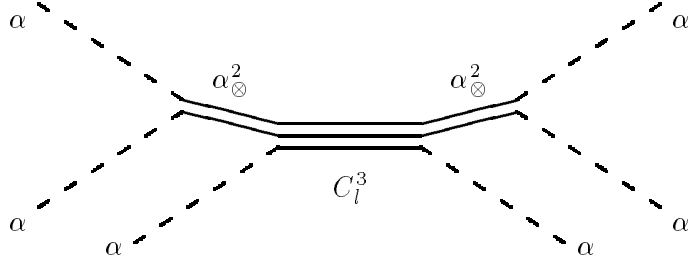

Figure 3: Stepwise fusion into the levels $C_l^3$

$$\epsilon_l - \epsilon_0 = \frac{1}{2\,\eta(S)}\left[\eta(\alpha_\otimes^3) - \eta(C_l^3)\right] \tag{5.8}$$

and ([7], (3.33))

$$\eta(\alpha_\otimes^3) = -12\frac{2\mu-1}{\mu-2}(2\mu^2 - 4\mu + 1)\eta(S) \tag{5.9}$$

we can calculate $\eta(C_l^3)$. For $l = 6$ the function $\eta(C_6^3)$ is an average

$$\eta(C_6^3) = (1-\rho)\eta(C_{6,I}^3) + \rho\eta(C_{6,II}^3), \qquad 0 < \rho < 1 \tag{5.10}$$

of the anomalous dimensions of the true qp-fields $C_{6,I}^3$ and $C_{6,II}^3$. We shall see that $\rho$ is extremely small (Fig. 6)

$$\rho(\mu) = \mathcal{O}\left(10^{-2}\right) \tag{5.11}$$

The collective fusion algorithms starts with a list of the possible partitions for $l = 6$ and maximal length $r = 3$

$$\left[\{6,0,0\},\{5,1,0\},\{4,2,0\},\{3,3,0\},\{4,1,1\},\{3,2,1\},\{2,2,2\}\right] \tag{5.12}$$

With this basis the constraints (2.21) admit two solutions

$$
C_1 = \begin{pmatrix} (\delta_2)_3 \\ -6(\delta_2+1)_2(\delta_2+5) \\ +15(\delta_2+2)(\delta_2+4)_2 \\ -10(\delta_2+3)_3 \\ 0 \\ 0 \\ 0 \end{pmatrix}, \quad C_2 = \begin{pmatrix} 0 \\ 0 \\ (\delta_2)_3 \\ -(\delta_2)_2(\delta_2+3) \\ -(\delta_2+1)^2(\delta_2+2) \\ +2(\delta_2+1)_3 \\ -(\delta_2+2)^2(\delta_2+3) \end{pmatrix}, \tag{5.13}
$$

At $\mathcal{O}(1)$ we have

$$
\left\langle : \prod_{i=1}^{3} \alpha(x_i) :: \prod_{j=1}^{3} \alpha(y_j) : \right\rangle = \sum_{\substack{\text{permutations} \\ \pi\{1,2,3\}}} \prod_{i=1}^{3} \left((x_i - y_{\pi(i)})^2\right)^{-\delta_2} \tag{5.14}
$$

From (5.14), (5.13) we derive

$$
\sum_{\text{partitions}} C_a(k_1, k_2, k_3) C_b(l_1, l_2, l_3) \left\langle : \prod_{i=1}^{3} (q \cdot \partial)^{k_i} \alpha(x) :: \prod_{j=1}^{3} (q' \cdot \partial)^{l_j} \alpha(y) : \right\rangle \tag{5.12}
$$
$$
= M_{ab} \left(q \cdot (x-y)\right)^6 \left(q' \cdot (x-y)\right)^6 \left((x-y)^2\right)^{-3\delta_2 - 12} + q^2, (q')^2, q \cdot q' \text{ terms} \tag{5.15}
$$

We remember that the two-point function of two equal-rank symmetric traceless tensor qp-fields with equal dimensions (which is true here at $\mathcal{O}(1)$) is unique up to normalization (see [5], (A.11), (A.13)). The coefficient $M_{ab}$ of the "maximally polarized term" given in (5.15) contains all the available information. We find

$$
M = 2^{12} 6! (\delta_2)_3 (\delta_2)_4 \begin{pmatrix} m_{11} & m_{12} \\ m_{21} & m_{22} \end{pmatrix} \tag{5.16}
$$

with

$$
\begin{array}{rcl}
m_{11} & = & (\delta_2+4)_2 \left(33\delta_2^3 + 339\delta_2^2 + 1146\delta_2 + 1260\right) \\
m_{12} & = & (\delta_2)_2 (\delta_2+4)_2 (2\delta_2+5) \\
m_{22} & = & \frac{1}{30}(\delta_2)_2 \left(11\delta_2^3 + 52\delta_2^2 + 77\delta_2 + 40\right)
\end{array} \tag{5.17}
$$

which at $\delta_2 = 2$ reduces to

$$
M = 2^{12} \, 4! \, 6! \, 7! \begin{pmatrix} 5172 & 54 \\ 54 & \frac{7}{3} \end{pmatrix} \tag{5.18}
$$

Next we extract the anomalous dimensions. At $\mathcal{O}(\frac{1}{N})$ we should obtain on the r.h.s. of (5.15)

$$
\left(q \cdot (x-y)\right)^6 \left(q' \cdot (x-y)\right)^6 \left((x-y)^2\right)^{-18 - \eta(C_6^3)} \tag{5.19}
$$

instead of as in (5.15)

$$\left(q \cdot (x-y)\right)^6 \left(q' \cdot (x-y)\right)^6 \left((x-y)^2\right)^{-18-3\eta(\alpha)} \tag{5.20}$$

Therefore we must look for

$$+ \left[3\eta(\alpha) - \eta(C_6^3)\right] \log(x-y)^2 \tag{5.21}$$

These terms are obtained from the radiative correction graphs $B_{ij}$ [5], Fig. 7, inserted into the 6 $\alpha$-point function (5.14). For the following discussion it is helpful for the reader to have [5], Fig. 7 and the subsequent equations in mind. Since we want to maintain the sum over all permutations of the $y_j$-coordinates, crossed graphs ought to be omitted ($B_{13}$, $B_{22}$) whereas crossing symmetric graphs ($B_{12}$, $B_{23}$) contribute only one half. The factor $\log(x-y)^2$ arises from

$$-\log u \quad \to \quad +2\log(x-y)^2 \tag{5.22}$$

Moreover, we can neglect all powers of $u$, since they contribute only to trace terms in (3.6) or to $q^2$, $(q')^2$, $q \cdot q'$ terms in (5.15), respectively. Thus our relevant information is contained in the coefficients

$$A_m = \frac{1}{m!\eta(S)}\left\{a_{0m}^{(11)}z_1^3 + a_{0m}^{(21)}z_1^2 + \frac{1}{2}a_{0m}^{(23)}z_1^2\right\} \tag{5.23}$$

([5], (5.8)-(5.13) and (C.1), (C.7), (C.11) and [4], (2.20), (2.21)). These coefficients are connected with the crossing symmetric coefficients $\phi_m^{(2)}$ (5.7) by

$$\frac{\phi_m^{(2)}}{m!} = A_m + \sum_{r=0}^{m}\binom{m+1}{r+1}(-1)^r A_r \tag{5.24}$$

Explicitly the $A_m$ are

$$A_m = \frac{2\mu(2\mu-3)}{(\mu-2)(\mu)_m}\left\{2(2\mu-3)(\mu-1)_m - (1+2\frac{m+1}{m+2})(\mu-2)_{m+1} + (m+1)!\right\} \tag{5.25}$$

The same technique which was applied to (5.14) to derive (5.15), namely Taylor expansion and subsequent projection on $C_1$, $C_2$, is now applied to

$$\sum_{\substack{\text{permutations}\\\pi\{1,2,3\}}} \prod_{i=1}^{3}\left((x_i - y_{\pi(i)})^2\right)^{-\delta_2} \sum_{m=0}^{6} A_m\Big[(1 - v_{12,\pi(1)\pi(2)})^m + \tag{5.26}$$
$$(1 - v_{13,\pi(1)\pi(3)})^m + (1 - v_{23,\pi(2)\pi(3)})^m\Big]$$

$$v_{ij,kl} = \frac{(x_i - y_l)^2(x_j - y_k)^2}{(x_i - y_k)^2(x_j - y_l)^2} \tag{5.27}$$

and yields a matrix $Q_{ab}$

$$Q = 2^{12}4!\,6!\,7!\begin{pmatrix} q_{11} & q_{12} \\ q_{21} & q_{22} \end{pmatrix} \tag{5.28}$$

$$q_{11} = \frac{6}{7} \frac{(2\mu - 3)}{(\mu - 2)(\mu + 1)_5} [\, 26602\,\mu^7 + 385729\,\mu^6 + 2061655\,\mu^5 + 4900671\,\mu^4 + \quad (5.29)$$
$$+ 4415855\,\mu^3 + 3085040\,\mu^2 - 4738432\,\mu + 15929760\,]$$

$$q_{12} = q_{21} = \frac{1}{7} \frac{(2\mu - 3)}{(\mu - 2)(\mu + 1)_5} [\, 626\,\mu^7 + 9077\,\mu^6 + 48515\,\mu^5 + 118155\,\mu^4 + \quad (5.30)$$
$$+ 129403\,\mu^3 + + 334768\,\mu^2 - 5504\,\mu + 997920\,]$$

$$q_{22} = \frac{1}{70} \frac{(2\mu - 3)}{(\mu - 2)(\mu + 1)_5} [\, 214\,\mu^7 + 3103\,\mu^6 + 16585\,\mu^5 + 40801\,\mu^4 + \quad (5.31)$$
$$+ 47921\,\mu^3 + 60816\,\mu^2 + 104960\,\mu + 431200\,]$$

The eigenvalues $\lambda_{I,II}$ obtained from

$$\det(Q - \lambda M) = 0 \quad (5.32)$$

correspond (due to (5.21) - (5.23)) to

$$\lambda_{I,II} = \frac{1}{2\,\eta(S)} \Big[ 3\eta(\alpha) - \eta(C^3_{6,I,II}) \Big] \quad (5.33)$$

and come out as

$$\lambda_{I,II} = \frac{2\mu - 3}{140(\mu - 2)(\mu + 1)_5} \Big[ f(\mu) \pm \mu(\mu - 1)\sqrt{3\,D(\mu)} \Big] \quad (5.34)$$

$$f(\mu) = 454\,\mu^7 + 6583\,\mu^6 + 35185\,\mu^5 + 84081\,\mu^4 + 79361\,\mu^3 + 42056\,\mu^2 -$$
$$- 12520\,\mu + 369600 \quad (5.35)$$

$$D(\mu) = 24652\,\mu^{10} + 764212\,\mu^9 + 10507915\,\mu^8 + 84605664\,\mu^7 +$$
$$+ 443501806\,\mu^6 + 1596002156\,\mu^5 + 4066215915\,\mu^4 + 7437372768\,\mu^3 +$$
$$+ 9600480112\,\mu^2 + 8084044800\,\mu + 4400120000 \quad (5.36)$$

In Fig. 4 both $\eta(C^3_{6,I})$ and $\eta(C^3_{6,II})$ are given. Since $D(\mu) > 0$ (no double zero) these curves are smooth. It turns out that $\eta(C^3_6)$ (5.10) is indistinguishable from $\eta(C^3_{6,I})$ in Fig. 4. However, we plot the difference $\eta(C^3_{6,I}) - \eta(C^3_6)$ in Fig. 5 on an enlarged scale, which shows that $\rho$ is positive as required, Fig. 6.

## 6 General classes of qp-fields

A general class $(Y, p)$ with Young frame $Y$ and block number $\#Y$ involves

$$n = \frac{1}{2}(p - \#Y) \in I\!N \quad (6.1)$$

contractions of vector indices. At leading order all levels in such class are accessible by normal products. Consequently there is an algorithm which allows to determine all qp-fields and their anomalous dimensions at leading order. This leading order is $\mathcal{O}(\frac{1}{N^n})$ for the fusion coefficients and $\mathcal{O}(\frac{1}{N^{n+1}})$ for the dimensions. The energy-momentum tensor with $n = 1$ is a typical example. We remind the reader that normal products

$$: \vec{S}\vec{S} : \quad \text{and} \quad : \vec{S}\alpha : \tag{6.2}$$

are not permitted (see [8] and references therein).

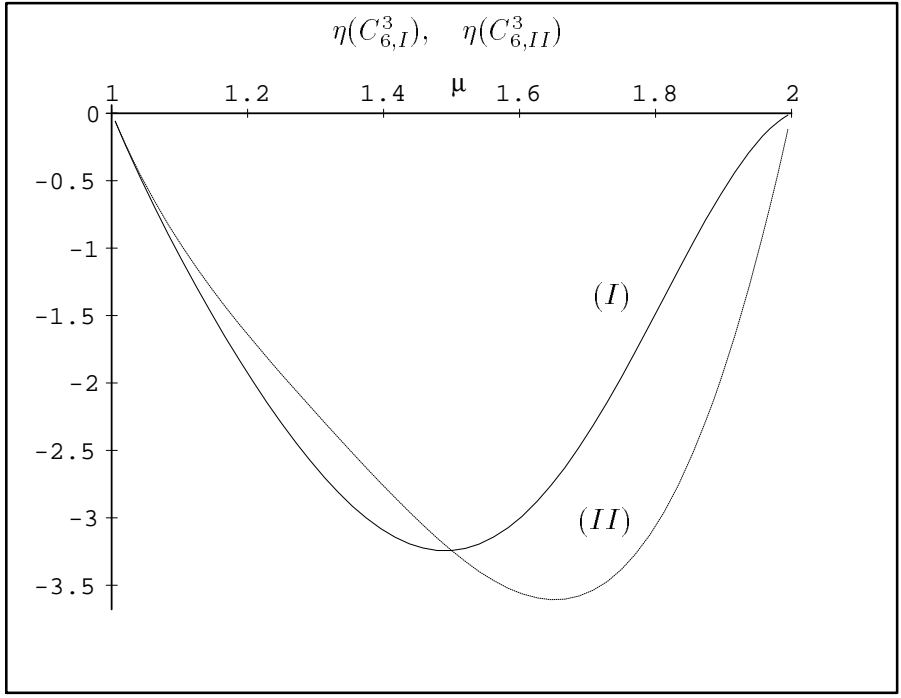

Figure 4: $\eta(C^3_{6,I})$, $\eta(C^3_{6,II})$ as functions of $\mu$

Let us now consider the simpler classes with

$$p = \#Y, \qquad n = 0 \tag{6.3}$$

The levels in such class are denoted

$$M_L^{pr} \qquad (L : \quad \text{tensor rank}) \tag{6.4}$$

If

$$e^Y_{a_1,a_2,\ldots,a_p;\, b_1,b_2,\ldots,b_p} \tag{6.5}$$

is the symmetrizer for the Young frame $Y$, we can construct a normal product for the level (6.4) as

$$\phi^{Y,L}_{\{k_1,\ldots,k_p;\, l_1,\ldots,l_r\}} = \sum_{\{b_i\}} e^Y_{a_1,\ldots,a_p;\, b_1,\ldots,b_p} : \prod_{i=1}^p \left[\left(\partial_{\mu_i}\right)^{k_i}_\otimes S_{b_i}\right] \prod_{j=1}^r \left[\left(\partial_{\nu_j}\right)^{l_j}_\otimes \alpha\right] : - \text{traces} \tag{6.6}$$

where

$$\sum_{i=1}^p k_i = k, \qquad \sum_{j=1}^r l_j = l, \qquad k + l = L \tag{6.7}$$

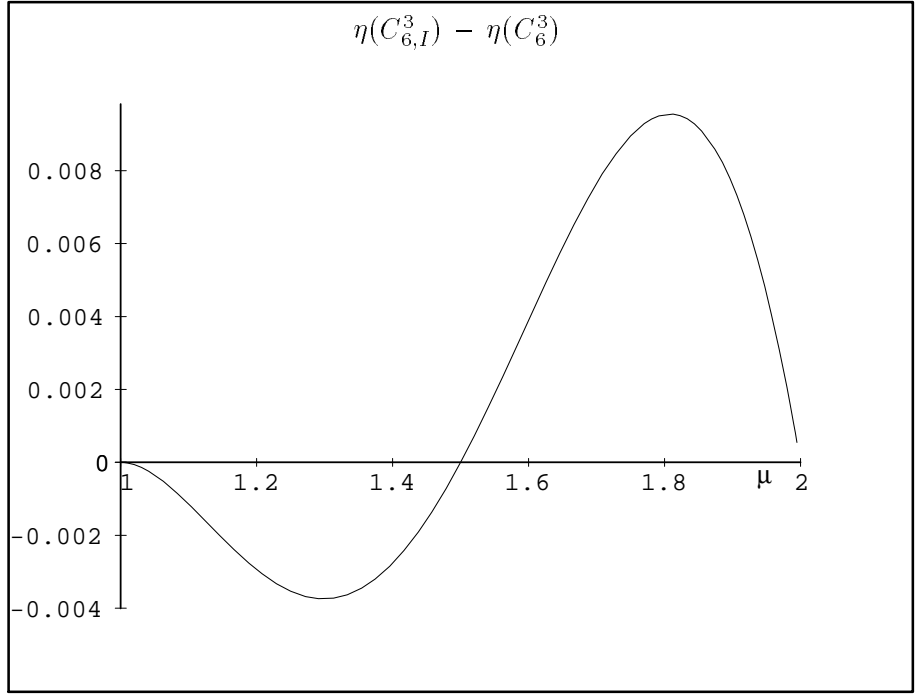

Figure 5: The difference $\eta(C^3_{6,I}) - \eta(C^3_6)$ as function of $\mu$

We can assume that

$$k_1 \geq k_2 \geq \ldots \geq k_p \geq 0$$
$$l_1 \geq l_2 \geq \ldots \geq l_r \geq 0 \qquad (6.8)$$

In order to construct a qp-field we consider linear combinations

$$\sum_{\substack{\text{partitions of} \\ L \text{ of type } (p,r)}} C_{\{k_1,\ldots,k_p;l_1,\ldots,l_r\}} \phi^{Y,L}_{\{k_1,\ldots,k_p;l_1,\ldots,l_r\}} \qquad (6.9)$$

where "type $(p,r)$" is defined by (6.7)-(6.8) with variable $k$ (or $l$). These linear combinations are submitted to the constraints

$$\sum_{\substack{\text{partitions of} \\ L \text{ of type } (p,r)}} C_{\{k_1,\ldots,k_p;l_1,\ldots,l_r\}} \left\{ \sum_{i=1}^{p} k_i(\delta_1 + k_i - 1) \phi^{Y,L-1}_{\Omega\{k_1,\ldots,k_i-1,\ldots,k_p;l_1,\ldots,l_r\}} + \right. \qquad (6.10)$$
$$\left. \sum_{j=1}^{r} l_j(\delta_2 + l_j - 1) \phi^{Y,L-1}_{\Omega\{k_1,\ldots,k_p;l_1,\ldots,l_j-1,\ldots,l_r\}} \right\}$$
$$(\delta_1 = dim(S), \quad \delta_2 = dim(\alpha))$$

where $\Omega$ as before reestablishes the order (6.8).

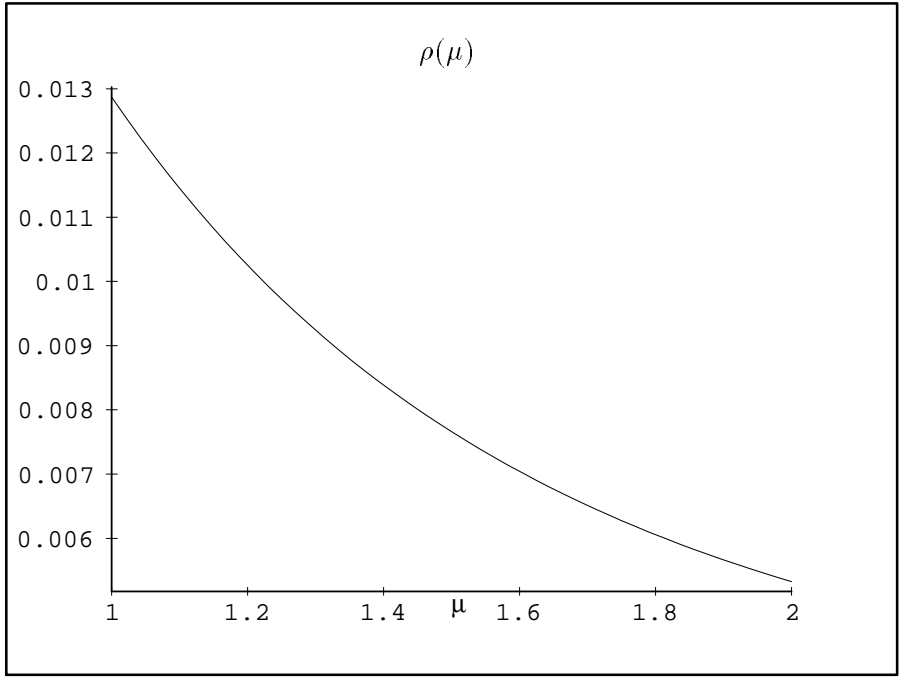

Figure 6: The function $\rho(\mu)$

In the case of Young frames with one row the number of qp-fields at such level (6.4) can easily be counted. Namely, for $p \cdot r \neq 0$ there are

$$\sum_{k=0}^{L} p_p(k) \, p_r(L-k) \tag{6.11}$$

fields $\phi^{Y,L}$ (6.6) and

$$\sum_{k=0}^{L-1} p_p(k) \, p_r(L-1-k) \tag{6.12}$$

constraints. Thus the degeneracy of the level $M_L^{p,r}$ is one for $L \in \{0,1\}$ and for $L = 2$ we have

$$\begin{array}{ll} 3, & \text{if } p \geq 2,\, r \geq 2 \\ 2, & \text{if } p = 1,\, r \geq 2 \quad \text{or} \quad p \geq 2,\, r = 1 \\ 1, & \text{if } p = r = 1 \end{array} \tag{6.13}$$

and in general the difference of (6.11) and (6.12). This difference gives a Hardy-Ramanujan distribution.

We consider now some examples. If $Y$ has one row only the levels $r = 0$ can be treated exactly as levels $p = 0$ in Section 3 and 4 (replace $\delta_2$ by $\delta_1$ in corresponding formulas). If $r = p = 1$ the level $L = 0$ is empty by (6.2) and all other levels are simply occupied (as found in [6], section 4).

Antisymmetric Young frames for O(N) imply also antisymmetric space-time tensors and counting of qp-fields becomes difficult. Consider the class

$$Y = \boxed{\phantom{x}}, \quad p = 2 \qquad (6.14)$$

and the tower $r = 0$. The level $L = 0$ is empty since the normal product vanishes identically. At $L = 1$ we get the Noether current

$$J_{\mu,ab} \;=\; : S_a \partial_\mu S_b - S_b \partial_\mu S_a : \qquad (6.15)$$

which is quasiprimary. For $L = 2$ we have two normal products

$$\begin{aligned} & : \partial_\mu \partial_\nu S_a S_b - S_a \partial_\mu \partial_\nu S_b : -\text{trace} \\ \text{and} \quad & : \partial_\mu S_a \partial_\nu S_b - \partial_\nu S_a \partial_\mu S_b : \end{aligned} \qquad (6.16)$$

Either one is a derivative of the current (6.15)

$$\begin{aligned} & \partial_\mu J_{\nu,ab} + \partial_\nu J_{\mu,ab} - \text{trace} \\ \text{respectively} \quad & \partial_\mu J_{\nu,ab} - \partial_\nu J_{\mu,ab} \end{aligned} \qquad (6.17)$$

So there is no qp-field at $L = 2$.

In this class and tower normal products are classified according to their partition $\{k_1, k_2\}$ and the Young frame for their space-time tensor which has at most two rows of lengths $[l_1, l_2]$, $(l_1 \geq l_2)$ so that

$$k_1 + k_2 = l_1 + l_2 = l \qquad (6.18)$$

For $l = 3$ we have normal products

$$\phi_{[3]\{3,0\}}, \quad \phi_{[3]\{2,1\}}, \quad \phi_{[2,1]\{2,1\}} \qquad (6.19)$$

and we form linear combinations with fixed tensor type $[l_1, l_2]$

$$C_{[3]\{3,0\}} \phi_{[3]\{3,0\}} + C_{[3]\{2,1\}} \phi_{[3]\{2,1\}}, \qquad C_{[2,1]\{2,1\}} \phi_{[2,1]\{2,1\}} \qquad (6.20)$$

At $l = 2$ we have

$$\phi_{[2]\{2,0\}} \quad \text{and} \quad \phi_{[1,1]\{1,1\}} \qquad (6.21)$$

and obtain two constraints for (6.20)

$$\begin{aligned} 3(\delta_1 + 2) C_{[3]\{3,0\}} + \delta_1 C_{[3]\{2,1\}} &= 0 \\ C_{[2,1]\{2,1\}} &= 0 \end{aligned} \qquad (6.22)$$

yielding a single qp-field. The rule for obtaining the representations $[l_1, l_2]$ from $\{k_1, k_2\}$ is to reduce two representations $[k_1]$ and $[k_2]$

$$[k_1] \otimes [k_2] \;=\; \sum [l_1, l_2] + \text{representations with contractions} \qquad (6.23)$$

It is obvious that a computer algebra approach to qp-fields and their dimensions is most convienient but also in reach, and therefore we want to cut the discussion short here. We want to conclude with the remark that the stepwise fusion procedure which we developed is not outdated by the collective fusion method discussed in this article, but that a simultanous treatment as in Section 4 is highly recommendable. If information about an infinite sequence of fields is desired it is still the only reasonable approach.

# References

[1] K. LANG AND W. RÜHL, *Z.Phys.* C **50** (1991) 285

[2] K. LANG AND W. RÜHL, *Z.Phys.* C **51** (1991) 127

[3] K. LANG AND W. RÜHL, *Nucl. Phys.* B **377** (1992) 371

[4] K. LANG AND W. RÜHL, *Phys. Letters* B **275** (1992) 93

[5] K. LANG AND W. RÜHL, *Nucl. Phys.* B **402** (1993) 573

[6] K. LANG AND W. RÜHL, *Nucl. Phys.* B **400** [FS] (1992) 597

[7] K. LANG AND W. RÜHL, Kaiserslautern University preprint KL-TH-93/7, *Z.Phys.* C, to appear

[8] K. LANG AND W. RÜHL, *Critical O(N) - vector nonlinear sigma-models: A résumé of their field structure*, Kaiserslautern University preprint, hep-th/9311046, (to appear in the Proceedings of the XXII Conference on Differential Methods in Theoretical Physics, Ixtapa, Mexico, September 1993)

[9] *Encyclopedic Dictionary of Mathematics*, 2nd ed., The MIT Press, Cambridge, Mass. and London, England 1968

[10] P. GODDARD, D. OLIVE, *Int. J. Mod. Phys.* **A1** (1986), 304